\documentclass[twocolumn,showpacs,preprintnumbers,amsmath,amssymb]{revtex4-1} 
\usepackage{graphicx,color}
\usepackage{dcolumn}
\usepackage{bm}
\usepackage{times}
\usepackage{subfigure}

\begin{document}

\title{Knotting of linear DNA in nano-slits and nano-channels: a numerical study}

\author{Enzo Orlandini}
\email{orlandini@pd.infn.it}
\affiliation{Dipartimento di Fisica e Astronomia and Sezione INFN,\\ Universit\`a di Padova,
Via Marzolo 8, 35131 Padova (Italy)
}

\author{Cristian Micheletti}
\email{michelet@sissa.it}
\affiliation{SISSA - Scuola Internazionale Superiore di Studi Avanzati and CNR-IOM Democritos,  Via Bonomea 265, 34136 Trieste (Italy)}

\date{\today}

\begin{abstract}
The amount and type of self-entanglement of DNA filaments is
signficantly affected by spatial confinement, which is ubiquitous in
biological systems.  Motivated by recent advancements in single DNA
molecule experiments based on nanofluidic devices, and by the
introduction of algorithms capable of detecting knots in open chains, we
investigate numerically the entanglement of linear, open DNA chains
confined inside nano-slits.  The results regard the abundance, type and
length of occurring knots and are compared with recent findings for DNA
inside nano-channels. In both cases, the width of the confining region,
$D$, spans the 30nm- 1$\mu$m range and the confined DNA chains are 1 to
4$\mu$m long. It is found that the knotting probability is maximum for
slit widths in the 70-100nm range. However, over the considered DNA
contour lengths, the maximum incidence of knots remains below 20\%,
while for channel confinement it tops 50\%. Further differences of the
entanglement are seen for the average contour length of the knotted
region which drops significantly below $D\sim 100$nm for
channel-confinement, while it stays approximately constant for slit-like
confinement. These properties ought to reverberate in different kinetic
properties of linear DNA depending on confinement and could be
detectable experimentally or exploitable in nano-technological
applications.  
\end{abstract}

\maketitle

\section{Introduction}\label{sec:intro}

Like other forms of entanglement, the abundance and type of knots in
equilibrated DNA molecules depends on both intrinsic and extrinsic
properties. The former include the chain contour length, the bending
rigidity and chirality while the latter include spatial constraints such
as confinement in narrow
spaces~\cite{micheletti_physrep,Marenduzzo_et_al_2010_JoP_Cond_Matt,muthukumar_adv_chem_phys}.

A notable example of spatially-confined DNA is offered by viral genome
packaged inside
capsids\cite{Marenduzzo_et_al_2010_JoP_Cond_Matt}. Typically, the viral
DNA contour length exceeds by orders of magnitude the capsid diameter,
resulting in a near crystalline density of the packaged
genome~\cite{Forrey:2006:Biophys-J:16617089,Leforestier:2009:Proc-Natl-Acad-Sci-U-S-A:19470490,Leforestier:2011:Biophys-J:21539789,Petrov:2007:J-Struct-Biol:17919923}. Over
the years, several numerical studies have accordingly tried to
understand not only how the DNA filament is packaged inside the virus
\cite{Kindt:2001:Proc-Natl-Acad-Sci-U-S-A:11707588,Arsuaga:2002:Biophys-Chem:12488021,Purohit:2003:Proc-Natl-Acad-Sci-U-S-A:12629206,Marenduzzo:2003:J-Mol-Biol:12842465,Arsuaga:2005:Proc-Natl-Acad-Sci-U-S-A:15958528,Forrey:2006:Biophys-J:16617089,Petrov:2007:J-Struct-Biol:17919923}
but especially how it can be ejected into the host cell through the
narrow capsid exit channel without being jammed by
self-entanglement~\cite{Arsuaga:2005:Proc-Natl-Acad-Sci-U-S-A:15958528,Matthews:2009:Phys-Rev-Lett:19257792,Marenduzzo:2009:Proc-Natl-Acad-Sci-U-S-A:20018693,rosa2012}. A
solution to this conundrum was proposed in
ref~\cite{Marenduzzo:2009:Proc-Natl-Acad-Sci-U-S-A:20018693} which
reported that the ordering effect of DNA cholesteric
self-interaction\cite{bouligand,Stanley:2005:Biophys-J} is responsible
for keeping the entanglement at a minimum and compatible with an
effective ejection process.

The effects of spatial constraints on DNA self-entanglement and the
possible implications on DNA condensation, packaging and translocation
have been systematically addressed only recently, largely because of the
introduction of suitable nano-devices and micro-manipulation techniques
that allow for probing the properties of few confined molecules
at a time~\cite{Tegenfeldt:2004:PNAS,Reisner:2005:Phys-Rev-Lett,Stein_et_al:2006:PNAS,Reisner:2007:Phys-Rev-Lett,Bonthuis:2008:Phys-Rev-Lett,doi:10.1021/mz300323a}.

In such contexts a still largely-unexplored research avenue is the
characterization of the occurrence of knots in open, linear DNA
molecules. In fact, theoretical and experimental studies of knot
occurrence have largely focused on equilibrated chains where knots are
trapped by a circularization reaction which ligates the two chain ends,
thus forming a ring.  The topology of such rings, is clearly maintained
until they open up, and therefore their knottedness is
well-defined.

This is not the case for open chains, where non-trivial entanglement
cannot be permanently trapped because of the two free ends. Yet, we are
all familiar with the fact that knots in open chains can be long-lived
and can affect various physical and dynamical properties of polymers. In
particular in lab-on-chip experiments the presence of knots in linear
DNAs may interfere with the confinement elongation process of the
molecules, an essential step for the detection of protein-DNA
interactions~\cite{Wang:2005:PNAS} and also towards genome sequencing by
pore-translocation~\cite{Zwolak_DiVentra:2008:RevModPhys,rosa2012}.
These considerations have stimulated a number of efforts aimed at suitably extending the
algorithmic notion of knottedness to linear, open chains~\cite{Millett:2005:Macromol,Min_entang_closure,Virnau:2005:J_Am_Chem_Soc}.

Building on these theoretical advancements and motivated by the upsurge of 
DNA nano-manipulation experiments~\cite{Tegenfeldt:2004:PNAS,Reisner:2005:Phys-Rev-Lett,Stein_et_al:2006:PNAS,Reisner:2007:Phys-Rev-Lett,Bonthuis:2008:Phys-Rev-Lett,doi:10.1021/mz300323a}
here we report on a numerical study of the knotting properties of linear DNA chains confined in nano-slits and nano-channels.

The investigation is based on a coarse-grained model of DNA and is a followup of two recent studies of the metric and entanglement properties
that we carried out for closed and open chains in nano-slits and
nano-channels~\cite{Micheletti:2012:Macromol,Micheletti:2012:SoftMatter}.
Specifically, the properties of knotted open chains confined in slits is reported here for the first time and is compared with the earlier
results for channel confinement.

\section{Methods}\label{sec:methods}

{\bf The model}

Hereafter we provide a brief, yet self-contained, description of the
coarse-grained DNA model and simulation and numerical techniques used to
characterize the topological properties in slit- and channel-like
confining geometries.

By following the approach of ref.~\cite{Rybenkov:1993:Proc-Natl-Acad-Sci-U-S-A:8506378}, linear
filaments of dsDNA are modeled as semi-flexible chains of $N$ identical cylinders.

The cylinders diameter is $d=2.5$nm, corresponding  to the dsDNA hydration diameter, and their axis length is equal to $b=10$nm, i.e. a fraction of the nominal dsDNA persistence length, $l_p=50$nm.

A chain configuration is fully specified by the location in space of the cylinders axes, $\vec{t}_1,\ \vec{t}_2, \vec{t}_N$ and, in the unconstrained case, its energy is given by the sum of two terms:

\begin{equation}
E=E_{excl-vol} + E_b \ .
\end{equation}

\noindent The first term accounts for the excluded volume interaction of the cylinders. It is equal to ``infinity''
if two non-consecutive cylinders overlap and zero otherwise. The second term is the bending energy, $E_b= - K_BT \frac{l_p}{b}\sum_{i} \vec t_i\cdot \vec t_{i+1}$ with $T=300$K being the system temperature and $K_B$ the Boltzmann constant. It is assumed that the DNA is in a solution with high concentration of monovalent counterions, so that its electrostatic self-repulsion is effectively screened. Because non-local self-contacts of the DNA molecule are infrequent for two- and one-dimensional confinement we neglect both desolvation effects and the DNA cholesteric interaction~\cite{Strey:1998:Curr-Opin-Struct-Biol:9666326,Marenduzzo:2009:Proc-Natl-Acad-Sci-U-S-A:20018693}. Finally, because linear chains can effectively relax torsion, the DNA torsional rigidity is neglected too.

The confinement of the DNA inside slits is enforced by requiring that the chain maximum span perpendicular to the slit plane
, $\Delta$, is lower than a preassigned value, $D$. Likewise, for channel confinement, it is required that the maximum calliper (diameter), $\Delta$,  measured perpendicularly to the channel axis is smaller than $D$. 

The conformational space of confined chains was explored by means of a Monte Carlo scheme employing standard local and global moves (crankshaft and pivot moves). Following the Metropolis criterion, a newly-generated trial conformation is accepted or rejected with probability given by $\min(1,\exp[-(E- \mu \Delta) / K_B T]$. In the latter expression $\mu$ represents an auxiliar parameter that couples to the chain span (or calliper size), $\Delta$. Accordingly, by using different values of $\mu$ it is possible to bias the sampling of the configurations towards configurations with different average values of $\Delta$. Next, because the biasing weight is set {\em a priori}, it is possible to remove it by using thermodynamic reweighting techniques and recover the canonical expectation values of the observables of interest. Advanced sampling and reweighting techniques (which are reviewed in detail in ref.~\cite{micheletti_physrep}) is adopted here because a direct enforcement of the geometrical constraints in the Monte Carlo sampling would be inefficient due to the high Metropolis rejection rate.
 
\begin{figure*}[htbp]
\includegraphics[width=6.0in]{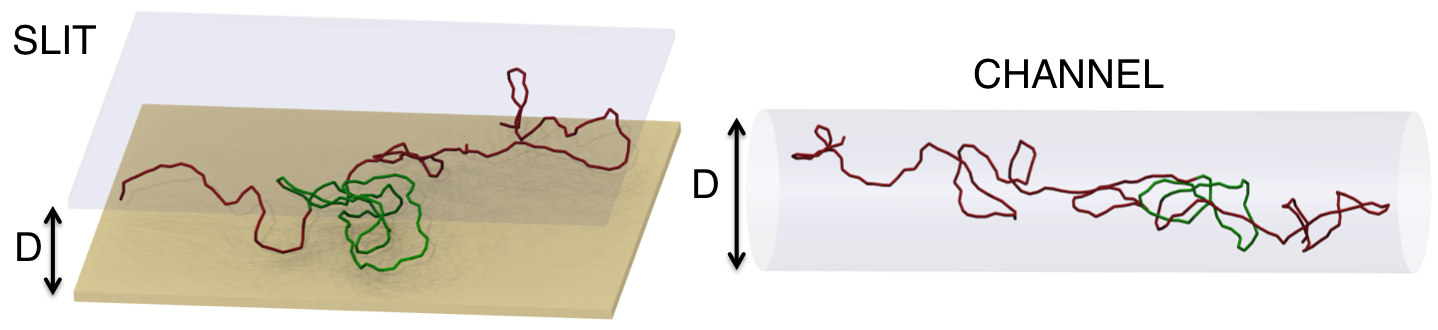}
\caption{Linear knotted chain of length $L_c=2.4\mu$m confined inside a slit and a channel. In both cases the confinement width is $D=70$nm and the chains accommodate a trefoil knot (highlighted in red).}
\label{fig:configurations}
\end{figure*}

We considered chains of $N=$120, 240, 360, 400 and $480$ cylinders, corresponding to contour lengths $L_c=Nb$ in the range 1--4.8$\mu$m. Across the various values of $\mu$ we collected $\sim 10^5$  uncorrelated configurations of chains that could be accommodated inside channels or slits with width $D$ in the $40$ nm--1$\mu$m range.  Notice that, because of excluded volume effects, the minimal width achievable by slits and channels in presence of a linear chain is $d$. 
It is therefore convenient to profile all properties in terms of the \emph{effective} width $D_{\rm eff} = D-d$.

{\bf Chain size and knot detection}
To characterize the average size of the chain we consider  its root mean square radius of gyration
\begin{equation}
R_g = \sqrt {\frac{1}{N} \langle \sum_{i=1}^N \sum_{\alpha=x,y,z}\left ( {\bf r}_{i,\alpha} - \bar{{\bf r}}_{\alpha}\right )^2\rangle },
\end{equation}
where $\alpha$ runs over the three Cartesian components of the position vector ${\bf r}_i$ of the center of the $i$-esim cylinder of the chain, $\bar{{\bf r}} = \frac{1}{N} \sum_{i=1}^N {\bf r}_i$ is the center of mass of the chain and the brackets $\langle \cdots \rangle$ denote the canonical 
average over chains in equilibrium an confined in slits or channels.

{\bf Profiling the knot spectrum}

The entanglement properties of the confined model DNA filaments were
characterised by establishing the knotted state of the open chains and
by measuring the knot contour length.

From a mathematical point of view only circular chains have a
well-defined topological knotted state since it cannot be altered by
distorting or changing the chain geometry as long as the chain
connectivity is preserved. To extend the concept of knottedness to open
chains, it is therefore necessary to close them in a ring with an auxiliar
arc~\cite{Millett:2005:Macromol,Min_entang_closure}. The knotted state
of the closed ring is then assigned to the open chain. The auxiliary arc
must clearly be suitably defined to ensure the robustness of the
topological assignment; in particular it must avoid interfering with the
self-entanglement of the open chain. To this purpose we adopted the
minimally-interfering closure scheme introduced in
ref.~\cite{Min_entang_closure}.

The position of the knot along the chain is next established by
identifying the shortest chain portion that, upon closure has the same
knotted state of the whole chain. To minimize the chance of detecting
slipknots~\cite{Millett:2005:Macromol} it is also required that the
complementary arc on the closed chain is unknotted.

Fig.~\ref{fig:configurations} illustrates two knotted configurations of 2.4$\mu$m-long open chains confined inside a slit and a channel. The knotted portion of the chains is highlighted.

\section{Results and discussion}

The metric properties of linear DNA chains for various  degrees of slit-like confinement were systematically addressed in ref.~\cite{Micheletti:2012:Macromol}. Such study indicated that for increasing confinement, the two principal axes of inertia of linear molecules first orient in the slit plane and next grow progressively as the chain spreads out in a quasi-two-dimensional geometry.

\begin{figure}[htbp]
\includegraphics[width=3.0in]{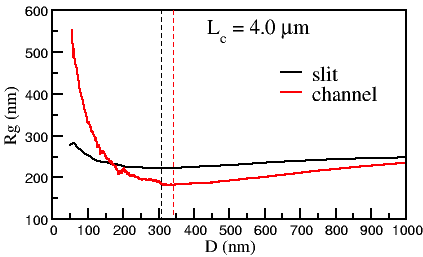}
\caption{Average radius of gyration, $R_g$, of linear DNA with contour
  length $L_c=4.0 \mu m$ inside a slit and a cylindrical channel of
  effective width $D_{\rm eff}$. Black and red dashed lines mark the location of the minimum of $R_g$ respectively for slits and
  channels.}
\label{fig:metric properties}
\end{figure}

The interplay of the increase of the chain size projected in the slit
plane, $R_{||}$, and the concomitant decrease of the transverse size,
$R_{\perp}$, results in the non-monotonic behaviour of $R_g$, as
illustrated in Fig.~\ref{fig:metric properties}. The data shows that $R_g$ attains a minimum at an effective channel width, $D^*$,
that is slightly larger then the average radius of gyration of the unconstrained, bulk case (marked by the black dashed line).

 For comparative purposes, in the same figure it is shown the average
 radius of gyration of equally-long linear DNA chains confined in
 channels. In this case too one observes the non-monotonic dependence of
 $R_g$ on the width of the confining region, which attains its minimum
 at a channel width, $D^*$ slightly larger that the slit case (red dashed line).
However, the increase of $R_g$
 past the minimum is much more dramatic than for the slit case. This
 reflects the fact that the chain can only elongate in only one
 direction rather than in a plane.

A relevant question regards the extent to which the dimensionality of
the confining region (slits or channels) and its width can affect the
incidence, complexity and size of knots in linear chains.
We accordingly applied the minimally-interfering closure scheme to
 establish the knotted state of equilibrated linear chains and
to locate the knot along their contour.

\begin{figure*}[htbp]
\includegraphics[width=6.0in]{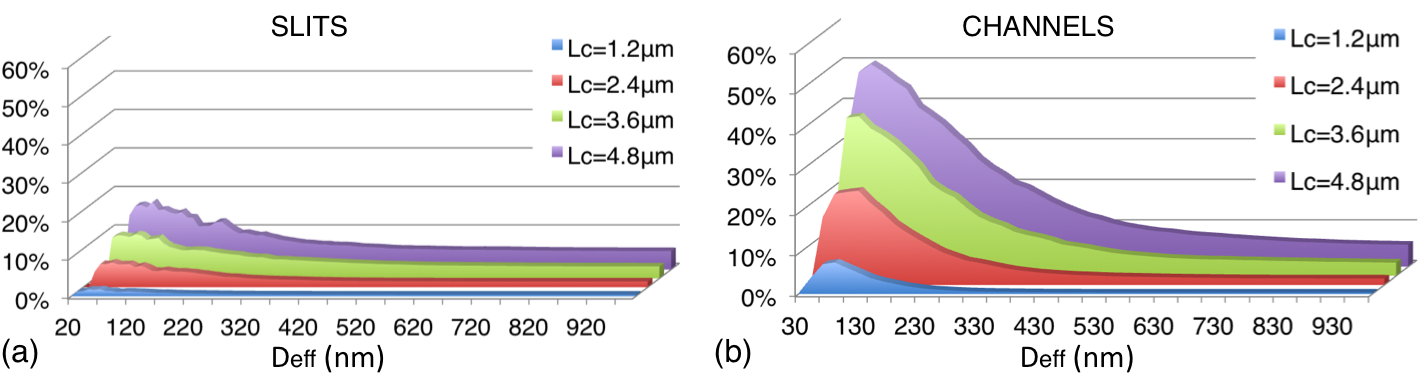}
\caption{Percentage of knotted linear DNA chains inside slits (a) and
  cylindrical channels (b) of transversal effective size $D_{\rm eff}$.
  Different curves refer to different DNA contour lengths. The estimated
  relative errors on the knotting probability are smaller than $1\%$ for
  $D_{\rm eff} >200$ nm and about of $3-4\%$ for $D_{\rm
    eff}<200$ nm. The knotting data for linear chains inside channels are based on the study of ref.~\cite{Micheletti:2012:SoftMatter}.}
\label{fig:knotting_probabilities}
\end{figure*}

We first discuss the overall incidence of non-trivial knot topologies.
The results for slit-confined chains are shown in
Fig.~\ref{fig:knotting_probabilities}(a).  As expected, for each fixed
value of $D_{\rm eff}$ the knotting probability depends strongly on
$L_c$~\cite{micheletti_physrep}.  In fact, going from $L_c=1\mu$m to
$4.8\mu$m the knotting probability increases by one order of
magnitude. By comparison, the knotting probability variations on $D_{\rm
  eff}$ (at fixed $L_c$) are smaller, though noticeable. More
importantly, the knotting probability displays, as $D_{\rm eff}$
decreases, a non-monotonic behaviour with a maximum enhancement peak at
a width, $D_e$ that falls within the 50-100 nm range.  

In particular, the knotting probability varies by a factor of $2$ going from the unconstrained case $D\sim 1\mu$m to $D_e \sim 80$nm. As consistently indicated by analogous results of slit-confined rings\cite{Micheletti:2012:Macromol}, this knotting enhancement should be measurable experimentally by circularizing dsDNA molecules with complementary single-stranded ends inside slits.

As shown in Fig.~\ref{fig:knotting_probabilities}(b), the confinement-induced enhancement 
of non-trivial knots is even stronger for the channels case: at
the largest contour length, $L_c = 4.8\mu$m, the probability peak value is
about 10 times larger than the bulk one. One may anticipate that the maximum knot enhancement is attained for the
same channel width, $D^*$, for which $R_g$ is minimum ( i.e.  at the highest
value of the overall chain density).  This is, however, not the case
since $D_e$ is about one third of $D^*$ at all considered chain lengths.

\begin{figure*}[htbp]
\includegraphics[width=6.0in]{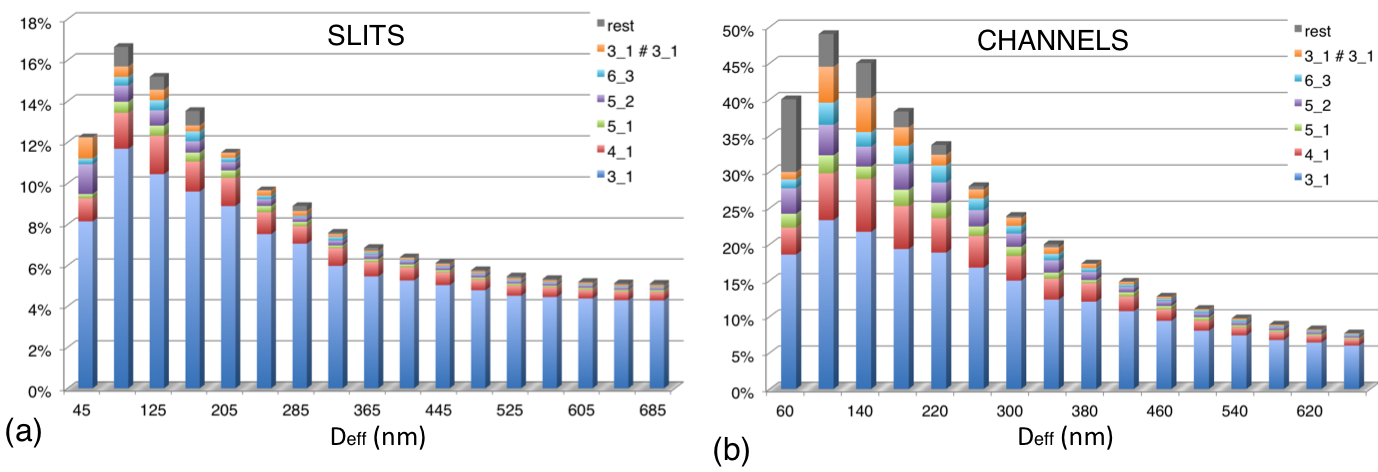}
\caption{Knot spectrum as a function of $D_{\rm eff}$ for 4.8$\mu$m-long linear DNA chains confined within slits (a) and channels (b).  The knotting data for linear chains inside channels are based on the study of ref.~\cite{Micheletti:2012:SoftMatter}.}
\label{fig:knot_spectrum}
\end{figure*}

Besides the overall knotting probability, it is of interest to examine in which proportion the various types of knots contribute to the overall knot population.   The results for 4.8$\mu$m-long chains inside slits are shown in Fig. ~\ref{fig:knot_spectrum}.
One notices that,  at all explored values of $D_{\rm eff}$, prime and composite knots with up to six crossings in their minimal diagrammatic representation account for at least 90\% of the observed knot population. In particular the simplest knot type, the  $3_1$ or trefoil knot, is by far the most abundant.  For channel confinement the predominance of  simple knots is even stronger. In particular the peak probability of the trefoil knot is, at $D_{\rm eff}=D_e$  about $23\%$, which is four times larger than for the unconstrained, bulk case.   

The result is noteworthy for several reasons. Firstly, at variance with the case of three-dimensional isotropic confinement (cavity, capsids) of DNA rings ~\cite{Micheletti:2008:Biophys-J:18621819}  the  knot spectrum of chains in slits and channels is dominated by the simplest knot types. This fact  was recently established for closed chains in slits and channels and for open chains in channels only~\cite{Micheletti:2012:Macromol,Micheletti:2012:SoftMatter}. The present results for open chains inside slits therefore complete the overall picture in a consistent way. Secondly, the percentage  of simple knots (and in particular of the simplest one, the trefoil)  found in two-dimensional confinement  (slits) is at least doubled in one-dimensional confinement (i.e. channels).

\begin{figure*}[floatfix]
(a)\includegraphics[width=3.0in]{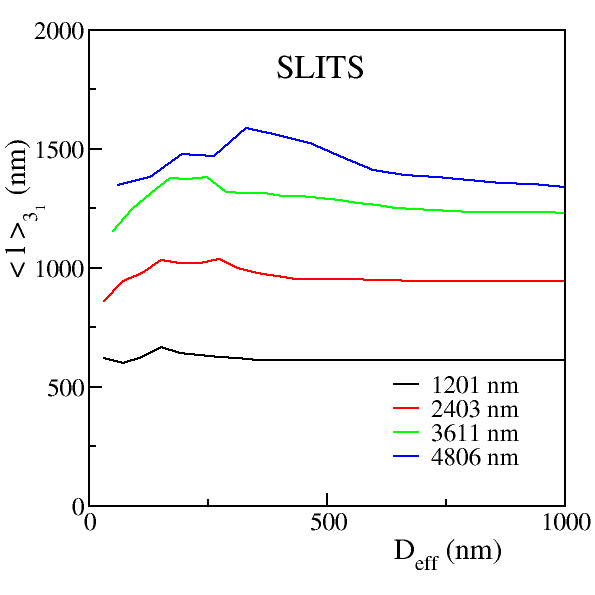}
(b)\includegraphics[width=3.0in]{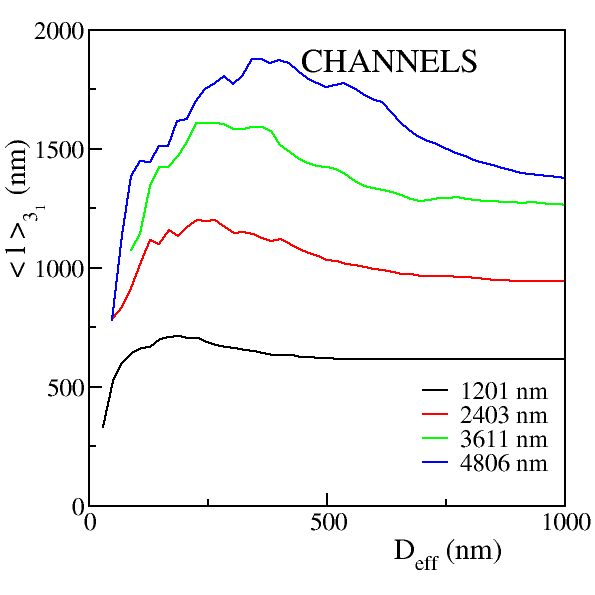}
(c)\includegraphics[width=3.0in]{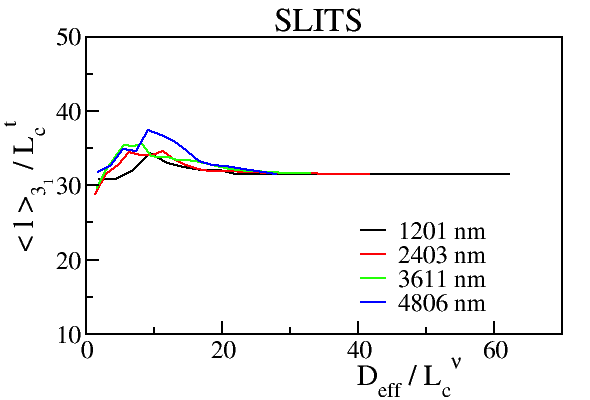}
(d)\includegraphics[width=3.0in]{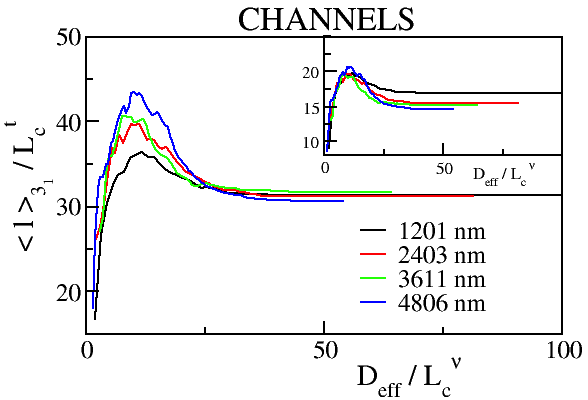}
\caption{Average contour length of knotted arcs  in linear DNA in slits (a) and channels (b) of  effective transverse dimension $D_{\rm eff}$.
In (c) and (b) the data of (a) and (b) are rescaled by $L_c^t$ for $l_{3_1}$ and by $L_c^{\nu}$ with $\nu=0.585$ for $D_{\rm eff}$.
The main  panels refer respectively to $t=0.62\pm0.05$ while for the plot in the inset $t=0.74\pm 0.01$.}
\label{fig:slab}
\end{figure*}

To profile the finer characteristics of the chain self-entanglement, we
finally analysed the average length of the chain portion that is spanned by the
highly abundant trefoil knots, $l_{3_1}$. The results shown in 
Fig.~\ref{fig:slab} show that, at fixed $L_c$, $l_{3_1}$ is non-monotonic on $D_{\rm
  eff}$ for both slits and channels.
 However, one major difference is that is that the knot length decreases dramatically after
the peak for channels while for slits it appears
to approach a limiting value not dissimilar from the bulk one. 

For both types of confinement, $l_{3_1}$ depends strongly on $L_c$. For instance, the confinement width associated
with the maximum knot length varies noticeably with $L_c$. In addition,
the relative difference of the peak value of $l_{3_1}$
respect to the bulk case increases with $L_c$ too (it ranges from 10\%
to 22\% for slits and from 16\% to 36\% for the channels).

It is interesting to examine the observed dependence of $l_{3_1}$ on
$L_c$ in connection with earlier scaling studies on closed rings in various conditions (unconstrained, collapsed, stretched etc.\cite{Farago:2002:EPL,Marcone:2005:J-Phys-A,Marcone:2007:PRE,Tubiana:2011:PRL}). In particular, we recall that knots
in unconstrained rings are known to be weakly localized in that their
contour length scales sublinearly with $L_c$.  Specifically, for
trefoils it was shown that for asymptotically-long rings, $l_{3_1} \sim
L_c ^t$ with $t \approx 0.65$~\cite{Marcone:2007:PRE}.

Motivated by these earlier findings we analysed the data in
Fig.~\ref{fig:slab} and rescaled them so to collapse the $l_{3_1}$ curves for very weak confinement ($D_{\rm eff} > 800 nm$). 
The optimal rescaling was obtained for the exponent, $t = 0.62 \pm 0.05$ for both slits and channels. For stronger confinement, the
rescaled curves depart from each other. By contrast with the slit case, the channel data show a systematic
upward trend for increasing  $L_c$. As a matter of fact, the portion of the curve for $D_{\rm eff}\lesssim D_e$ shows a
good collapse for $t=0.74\pm 0.01$.

This suggests that, as confinement increases, the knotted subregion of
the chain remains weakly localized but it further swells along the
unconstrained dimensions. 

\section{Conclusions}\label{sec:conclusions}

In polymer physics there is an ongoing effort to understand the
extent to which spatial constraints affect the probability of
occurrence, the complexity and size of topological defects in linear
polymers. For DNA this problem has various implications both for the understanding of some
biological elementary processes (such as translocation and viral ejection)
and for the development of efficient setups for DNA nano-manipulation protocols (such as sorting or sequencing).
 
Here we reported on a numerical study, based on advanced Monte Carlo
simulations, thermodynamic reweighting and scaling analysis of the
equilibrium topological properties of a coarse grained model of linear
DNA confined in rectangular slits and cylindrical nanochannels. The
investigation was carried out for linear chains with contour lengths
ranging between $1.2$ and $4.8 \mu$m and confined within geometries
whose transversal dimension $D_{\rm eff}$ span continuously the
$30-1000$nm range.

We found that, both for slits and channels, the knotting probability is
a non monotonic function of $D_{\rm eff}$ with a peak that occurs at a
length-dependent confinement width $D_e$. Most importantly and unlike DNA in capsids
(i.e. under full confinement) the enhancement of the topological
entanglement in slits and channels is not followed by a corresponding
enhancement of the entanglement complexity.  Indeed, despite that the
peak knotting probability exceeds by several times the one in
the bulk, most of the knots observed belong to very simple knot types.
This effect is particularly evident for channel confinement. This
suggests that nano-fluidic devices based on this or similar one-dimensional
geometries may be very effective for producing a good population of
linear DNA molecules with a simple knot tied in.

Finally, by using a robust algorithm for locating knots
in open chains~\cite{Min_entang_closure}, we show that the typical contour
length of the knotted region displays a non monotonic behaviour similar
to the one observed for the knotting probability.  Moreover, by looking
at its scaling behaviour as a function of the chain contour length, it is
found that for the whole range of confinement and both for slits and
channels, knots are weakly localized.

\begin{acknowledgements}
We thank L. Tubiana for useful discussions. We acknowledge support from
the Italian Ministry of Education, grant PRIN 2010HXAW77.
\end{acknowledgements}

\section*{Note}
The final version of this manuscript will be available at http://link.springer.com  as part of a special issue of the Journal of Biological Physics, DOI: 10.1007/s10867-013-9305-0.

\bibliographystyle{spmpsci}

\end{document}